\documentstyle[12pt,epsf]{article}
 \newcommand{\insertplot}[5]{\begin{figure}
 \hfill\hbox to 0.05in{\vbox to #5in{\vfill
 \inputplot{#1}{#4}{#5}}\hfill}
 \hfill\vspace{-.1in}
 \caption{#2}\label{#3}
 \end{figure}}
 \newcommand{\inputplot}[3]{
 \special{ps: plotfile #1}
}
\newcommand{\vphi}{\varphi}

\begin{document}

\title{
GRAVITATIONALLY BOUND MONOPOLES}
\vspace{1.5truecm}
\author{
{\bf Betti Hartmann}\\
Fachbereich Physik, Universit\"at Oldenburg, Postfach 2503\\
D-26111 Oldenburg, Germany\\
{\bf Burkhard Kleihaus}\\
Department of Mathematical Physics, University College, Dublin,\\
Belfield, Dublin 4, Ireland\\
{\bf Jutta Kunz}\\
Fachbereich Physik, Universit\"at Oldenburg, Postfach 2503\\
D-26111 Oldenburg, Germany}

\vspace{1.5truecm}

\date{\today}

\maketitle
\vspace{1.0truecm}

\begin{abstract}
We construct monopole solutions in SU(2) Einstein-Yang-Mills-Higgs theory,
carrying magnetic charge $n$. For vanishing and small Higgs selfcoupling, 
these multimonopole solutions are gravitationally bound.
Their mass {\it per unit charge} is lower than the mass of the
$n=1$ monopole. For large Higgs selfcoupling only a repulsive phase exists.
\end{abstract}
\vfill
\noindent {Preprint hep-th/0009195} \hfill\break
\vfill\eject

\section{Introduction}

Magnetic monopoles arise as topological defects in theories
which undergo spontaneous symmetry breaking.
In general, magnetic monopoles exist if the mapping
of the vacuum manifold, associated
with the symmetry breaking, onto the two-sphere is non-trivial.
The existence of magnetic monopoles is consequently a generic prediction
of grand unification.
If magnetic monopoles were indeed present in the universe,
they would have a host of astrophysical and cosmological 
consequences.

Here we consider magnetic monopoles
\cite{mono} and multimonopoles \cite{multi,rr,kkt,sut} in
SU(2) Yang-Mills-Higgs (YMH) theory, with the Higgs field in the
triplet representation. 
These solutions are globally regular.
Since their magnetic charge is proportional to their topological charge, 
the regular monopole and multimonopole solutions reside in
topologically non-trivial sectors of the theory.
In contrast, the monopole-antimonopole pair solutions \cite{taubes,map}
are topologically trivial.

In the  Bogomol'nyi-Prasad-Sommerfield (BPS) limit \cite{B,PS}, 
where the strength of the Higgs self-interaction potential vanishes,
the mass of the monopole and multimonopole solutions saturates
its lower bound, the Bogomol'nyi bound. In particular,
the mass {\it per unit charge} of an $n>1$ monopole is precisely
equal to that of the $n=1$ monopole.
There is no interaction between the monopoles in the BPS limit,
because the spatial components of the stress tensor vanish \cite{JT}.
The Higgs field is massless and mediates a
long range attractive force which exactly cancels the long range repulsive
magnetic force of the $U(1)$ field
\cite{M,N}.

For finite Higgs selfcoupling, however,
the Higgs field is massive and therefore decays exponentially.
Consequently the long range magnetic field dominates at large distances,
leading to the repulsion of like monopoles \cite{Gold}.
In particular, 
as verified numerically for $n=2$ and $n=3$ monopoles \cite{kkt},
the mass {\it per unit charge} 
of an $n>1$ monopole is higher than the mass of the $n=1$ monopole.
Thus for finite Higgs selfcoupling
there is only a repulsive phase between like monopoles.

Let us now consider the effect of gravity on the monopole solutions.
When gravity is coupled to YMH theory,
a branch of gravitating $n=1$ monopole solutions
emerges smoothly from the flat space monopole solution
\cite{gmono}.
The branch extends up to some maximal value of the
gravitational strength, beyond which
the size of the monopole core would be smaller than
the Schwarzschild radius of the solution
\cite{gmono}.
Along the branch with increasing gravitational strength, 
the mass of the gravitating monopole solutions decreases monotonically.
Finally a degenerate horizon forms \cite{gmono,lw,foot1,foot2}.

Similarly, in the presence of gravity
a branch of gravitating monopole-antimonopole pair (MAP) solutions
emerges from the flat space MAP solution \cite{map2}.

In this letter we investigate how gravity affects 
the static axially symmetric multimonopole solutions
of SU(2) YMH theory \cite{multi,rr,kkt}
and whether the inclusion of gravity
allows for an attractive phase of like monopoles.
For a given topological charge $n$, we find a branch of
globally regular and asymptotically flat multimonopole solutions,
emerging smoothly from the corresponding flat space solution,
and extending up to some maximal value of the gravitational strength.
Along the branch the mass of the solutions decreases monotonically.
And, indeed, we find a region of parameter space,
where the mass {\it per unit charge} 
of the gravitating multimonopole solutions is lower than 
the mass of the gravitating $n=1$ monopole.
Here the multimonopole solutions are gravitationally bound.
Thus gravity allows for an attractive phase of like monopoles,
not present in flat space.

\section{\bf Ansatz}

We consider SU(2) EYMH theory with action
\begin{equation}
S=\int \left ( \frac{R}{16\pi G} 
- \frac{1}{2e^2} {\rm Tr} (F_{\mu\nu} F^{\mu\nu})
-\frac{1}{4}{\rm Tr}(D_\mu \Phi D^\mu \Phi)
-\frac{1}{4}\lambda{\rm Tr}(\Phi^2 - \eta^2)^2
  \right ) \sqrt{-g} d^4x
\ , \end{equation}
with Newton's constant $G$, Yang-Mills coupling constant $e$, 
and Higgs self-coupling constant $\lambda$.

In flat space SU(2) YMH possesses multimonopole solutions
with axial symmetry for any topological number \cite{multi}.
For topological number $n \ge 3$, solutions with less symmetry
exist \cite{sut}. Here we extend the
axially symmetric multimonopole solutions to curved space.
We choose isotropic coordinates with metric \cite{kk}
\begin{equation}
ds^2=
  - f dt^2 +  \frac{m}{f} \left( d r^2+ r^2d\theta^2 \right)
           +  \frac{l}{f} r^2\sin^2\theta d\vphi^2
\ , \label{metric} \end{equation}
where $f$, $m$ and $l$ are only functions of $r$ and $\theta$.
The ansatz for the purely magnetic gauge field is \cite{rr,kk}
\begin{equation}
A_\mu dx^\mu =
\frac{1}{2er} \left[ \tau^n_\phi 
 \left( H_1 dr + \left(1-H_2\right) r d\theta \right)
 -n \left( \tau^n_r H_3 + \tau^n_\theta \left(1-H_4\right) \right)
  r \sin \theta d\phi \right]
\ , \label{gf1} \end{equation}
and for the Higgs field the ansatz is \cite{rr,kkt}
\begin{equation}
\Phi= \left(\Phi_1 \tau_r^{n}+\Phi_2 \tau_\theta^{n}\right)
\ . \end{equation}
The symbols $\tau^n_r$, $\tau^n_\theta$ and $\tau^n_\phi$
denote the dot products of the cartesian vector
of Pauli matrices, $\vec \tau = ( \tau_x, \tau_y, \tau_z) $,
with the spatial unit vectors
\begin{eqnarray}
\vec e_r^{\, n}      &=& 
(\sin \theta \cos n \phi, \sin \theta \sin n \phi, \cos \theta)
\ , \nonumber \\
\vec e_\theta^{\, n} &=& 
(\cos \theta \cos n \phi, \cos \theta \sin n \phi,-\sin \theta)
\ , \nonumber \\
\vec e_\phi^{\, n}   &=& (-\sin n \phi, \cos n \phi,0) 
\ , \label{rtp} \end{eqnarray}
respectively.
The winding number $n$ represents the topological charge of the solutions.
The four gauge field functions $H_i$ 
and the two Higgs field function $\Phi_i$ depend only on 
the coordinates $r$ and $\theta$.
For $n=1$ and $H_1=H_3=\Phi_2=0$, $H_2=H_4=K(r)$ and $\Phi_1=H(r)$,
the spherically symmetric ansatz of ref.~\cite{gmono} 
is recovered.

We fix the residual gauge degree of freedom \cite{rr,kk} 
by choosing the gauge condition 
$r\partial_r H_1-\partial_\theta H_2 =0$ \cite{kk}.

For asymptotically flat magnetically charged solutions
the boundary conditions at infinity are
\begin{equation}
f=m=l=1 \ , \ \ H_1=H_2=H_3=H_4=0 \ , \ \ \Phi_1=\eta \ , \Phi_2=0 
\ . \label{bc1a} \end{equation}
Requiring the solutions to be regular at the origin
($r=0$) leads to the boundary conditions
\begin{equation}
\partial_r f= \partial_r m= \partial_r l= 0 \ , \ \
 H_1=H_3 = 0 \ , \ \ H_2=H_4=1 \ , \ \Phi_1= \Phi_2=0 
\ . \label{bc2a} \end{equation}
The boundary conditions along the $\rho$- and $z$-axis
are determined by the symmetries of the solutions.
On both axes the functions 
$H_1$, $H_3$ and $\Phi_2$ and the derivatives
$\partial_\theta f$, $\partial_\theta m$, $\partial_\theta l$ 
$\partial_\theta H_2$, $\partial_\theta H_4$,
$\partial_\theta \Phi_1$
have to vanish.

Introducing the dimensionless coordinate 
$x=r\eta e$ and the Higgs field $\phi = \Phi/\eta$,
the equations depend only on two coupling constants, $\alpha$ and $\beta$,
\begin{equation}
\alpha^2 = 4\pi G\eta^2 \ , \ \ \ \beta^2 = \frac{\lambda}{e^{2}}
\ . \end{equation}

The mass $M$ of the multimonopole 
solutions can be obtained directly from
the total energy-momentum ``tensor'' $\tau^{\mu\nu}$
of matter and gravitation,
$M=\int \tau^{00} d^3r$ \cite{wein},
or equivalently from
$ M = - \int \left( 2 T_0^{\ 0} - T_\mu^{\ \mu} \right)
   \sqrt{-g} dr d\theta d\phi $,
yielding the dimensionless mass $\mu = \frac{e}{4\pi\eta} M$.

\section{\bf Results}

Subject to the above boundary conditions,
we solve the equations numerically \cite{xgmon}.
We start from the flat space (multi)monopole solutions \cite{kkt}
and increase $\alpha$.
Then branches of gravitating (multi)monopole solutions 
extend up to maximal values of $\alpha$, $\alpha_{\rm max}^{(n)}(\beta)$,
which increase with increasing topological charge
and decrease with increasing strength of the Higgs selfcoupling.

As the globally regular (multi)monopole solutions approach the critical value
of $\alpha$, 
a degenerate horizon starts to form \cite{gmono,lw,foot1}.
Indeed, the exterior of the critical solution corresponds to the exterior of
an extremal Reissner-Nordstr\o m (RN)
black hole with magnetic charge $n$ \cite{gmono,lw,foot2}.

In Fig.~1 we show the metric function $f$ 
of the gravitating $n=2$ multimonopoles in the BPS limit
for several values of $\alpha$.
With increasing $\alpha$ the function $f$ decreases monotonically
and approaches the metric function 
$f^{\rm RN}=\left(\frac{x/\alpha}{n+x/\alpha}\right)^2$ of the
extremal RN solution with magnetic charge $n=2$.
At $\alpha=1.5$, its value at the origin is already very small,
close to the value $f^{\rm RN}(0)=0$, representing the degenerate horizon
of the extremal RN solution. Also, the angle-dependence 
of the metric function is seen to diminish and disappear, 
as the critical value of $\alpha$ is approached.
Likewise, the Higgs field function $\phi_1$,
shown in Fig.~2, approaches the value $\phi_1^{\rm RN} = 1$
of the RN solution \cite{foot3}.
The Higgs field function $\phi_2$
and the gauge field functions approach their respective
RN values as well.
Satisfying Israel's theorem, the RN solution is spherically symmetric.

The mass {\it per unit charge}
of the (multi)monopole solutions decreases with increasing $\alpha$.
In the BPS limit, for $\alpha=0$ 
the mass {\it per unit charge} is precisely
equal to the mass of the $n=1$ monopole.
For $\alpha > 0$, however, we observe that
the mass {\it per unit charge} of the multimonopoles
is smaller than the mass of the $n=1$ monopole.
In particular, the mass {\it per unit charge} decreases
with increasing $n$.
The mass {\it per unit charge} of $n=2$ and 3 multimonopoles
in the BPS limit is shown in Fig.~3.
Thus in the BPS limit, there is an attractive phase between like monopoles,
not present in flat space.
Moreover, multimonopoles exist for gravitational coupling strength, 
too large for $n=1$ monopoles to exist.

For finite Higgs selfcoupling, the flat space multimonopoles
have higher mass {\it per unit charge} than
the $n=1$ monopole, allowing only for a repulsive phase
between like monopoles.
By continuity, this repulsive phase persists in the presence of gravity
for small values of $\alpha$, but it
can give way to an attractive phase for larger values of $\alpha$.
Thus the repulsion between like monopoles
can be overcome for small Higgs selfcoupling
by sufficiently strong gravitational attraction.
At the equilibrium value $\alpha_{\rm eq}$,
multimonopole mass {\it per unit charge} und monopole mass
equal one another.
We show the equilibrium value $\alpha_{\rm eq}$ in Fig.~4.
$\alpha_{\rm eq}$ increases with increasing Higgs selfcoupling,
yielding a decreasing region in parameter space for the attractive phase.
Finally, for large Higgs selfcoupling, only a repulsive phase is left.

While singly charged monopole solutions are stable,
stability of the static axially symmetric multimonopole solutions is not
obvious.
We conjecture, that the $n=2$ multimonopole solutions are stable,
as long as their mass {\it per unit charge} is lower than
the mass of the $n=1$ monopole.
For topological number $n \ge 3$, however, solutions with only
discrete symmetry exist in flat space \cite{sut}, which,
by continuity, should also be present in curved space
(at least for small gravitational strength).
For a given topological number $n>2$,
such multimonopole solutions without rotational symmetry
may possess a lower mass than the corresponding axially symmetric solutions.
The axially symmetric solutions may therefore not represent
global minima in their respective topological sectors,
even if their mass {\it per unit charge} is lower than
the mass of the $n=1$ monopole.

\section{\bf Black Holes}

Besides embedded abelian black hole solutions, like e.g.~the magnetically
charged RN solutions encountered above,
SU(2) EYMH theory also possesses genuine non-abelian black hole solutions
\cite{gmono}.
The SU(2) EYMH black hole solutions then are no longer uniquely
determined by their mass and charge (for vanishing angular momentum).
The non-abelian black hole solutions,
considered as black holes within (multi)monopoles, therefore
represent counterexamples to the ``no-hair'' theorem \cite{gmono}.

While static spherically symmetric ($n=1$) EYMH black holes were studied
in much detail \cite{gmono},
non-abelian black hole solutions with magnetic charge
$n>1$ were previously only obtained perturbatively \cite{ewein}.
Replacing the boundary conditions at the origin 
(\ref{bc2a}) by the appropriate
set of boundary conditions at some horizon radius $x_h$ \cite{kk,xgmon},
we obtain static axially symmetric black hole solutions with magnetic charge
$n>1$ numerically \cite{xgmon}.
These black hole solutions are asymptotically flat,
as required by the boundary conditions at infinity
(\ref{bc1a}), and they possess a regular deformed horizon \cite{kk,xgmon}.
Being static and not spherically symmetric,
these black hole solutions represent
new counterexamples to Israel's theorem.
While previous (non-perturbative) counterexamples \cite{kk,map3} 
to Israel's theorem are not classically stable,
black holes within multimonopoles
should provide classically stable counterexamples \cite{ewein}.

Finally we note, that the gravitational properties
of monopole spacetimes near the black hole threshold,
lend themselves to obtain new insight on the laws
of black hole thermodynamics \cite{lw2}.
Also, insight concerning the cosmic censorship hypothesis
might be obtained from the static black hole solutions
without spherical symmetry \cite{foot4}.

{\bf Acknowledgement}

We would like to thank the RRZN in Hannover for computing time.

\newpage

\begin{figure}\centering\epsfysize=8cm
\mbox{\epsffile{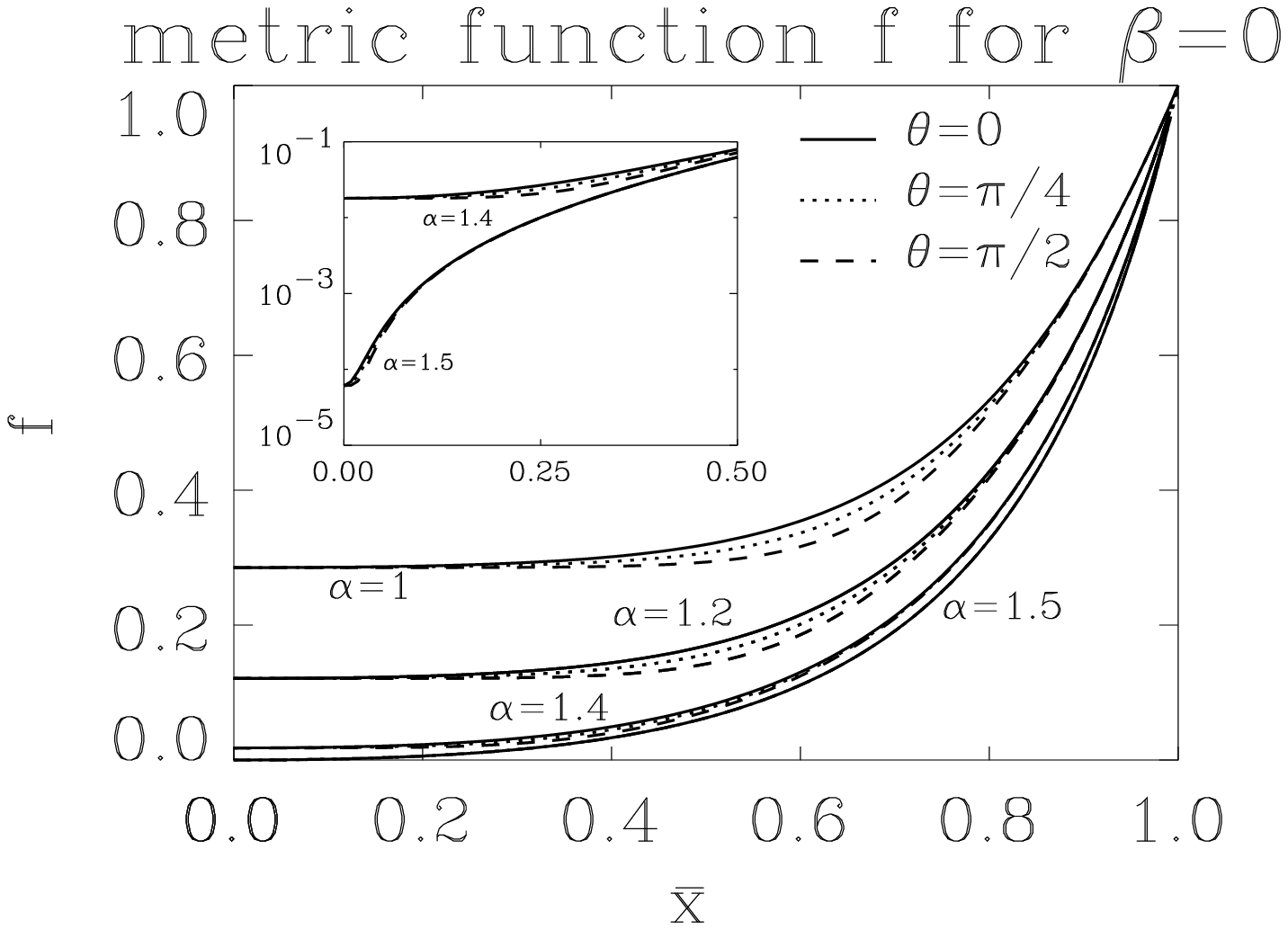}}
\caption{
The metric function $f$ is shown as a function of the compactified
dimensionless radial coordinate $\bar x = x/(1+x)$
for three values of the angle $\theta$
for the $n=2$ multimonopole solutions
in the BPS limit, for a sequence of values of $\alpha$,
approaching the critical value $\alpha_{\rm cr}$.
}
\end{figure}

\begin{figure}\centering\epsfysize=8cm
\mbox{\epsffile{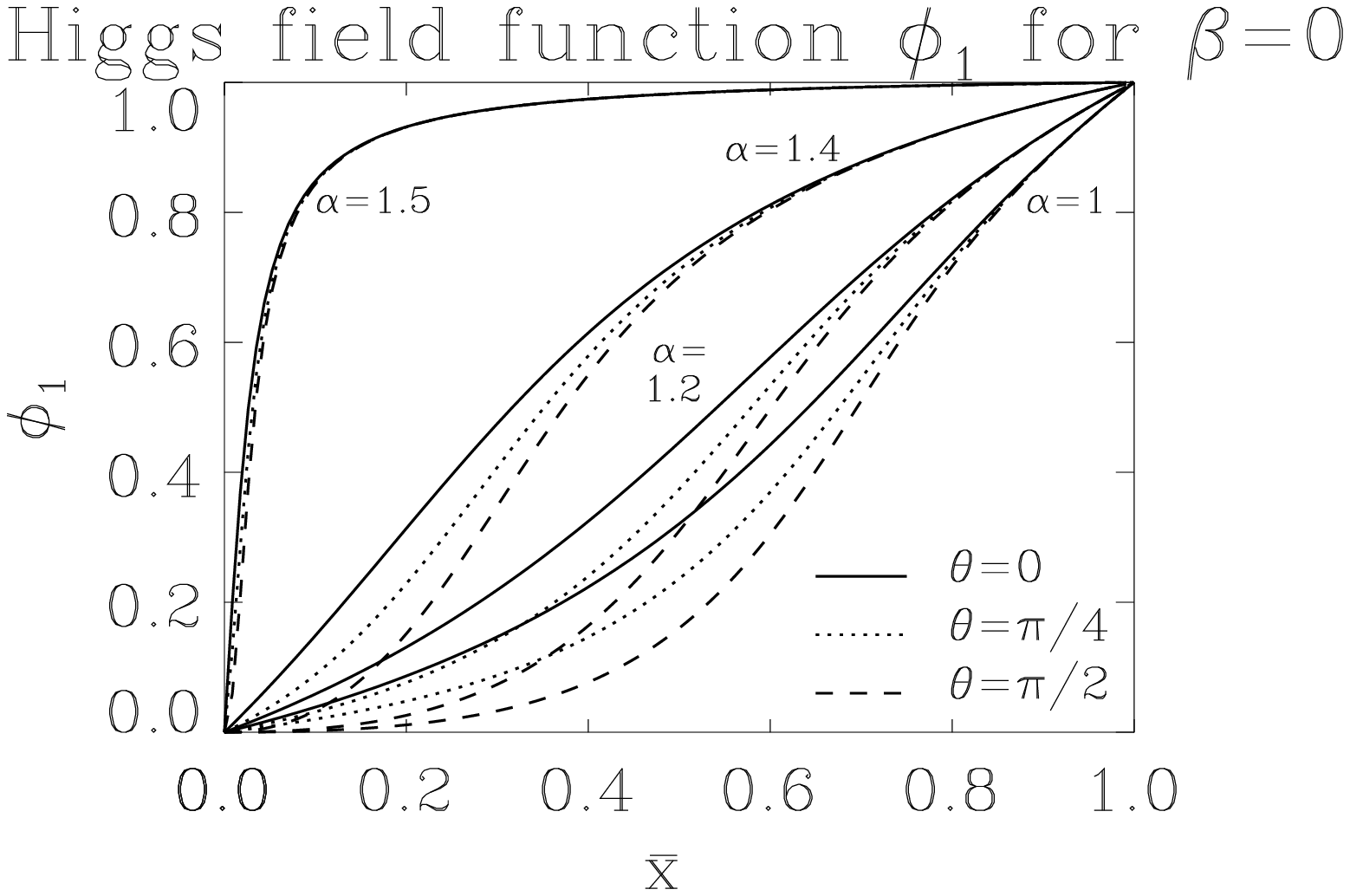}}
\caption{
Same as Fig.~1 for the Higgs field function $\phi_1$.
}
\end{figure}

\begin{figure}\centering\epsfysize=8cm
\mbox{\epsffile{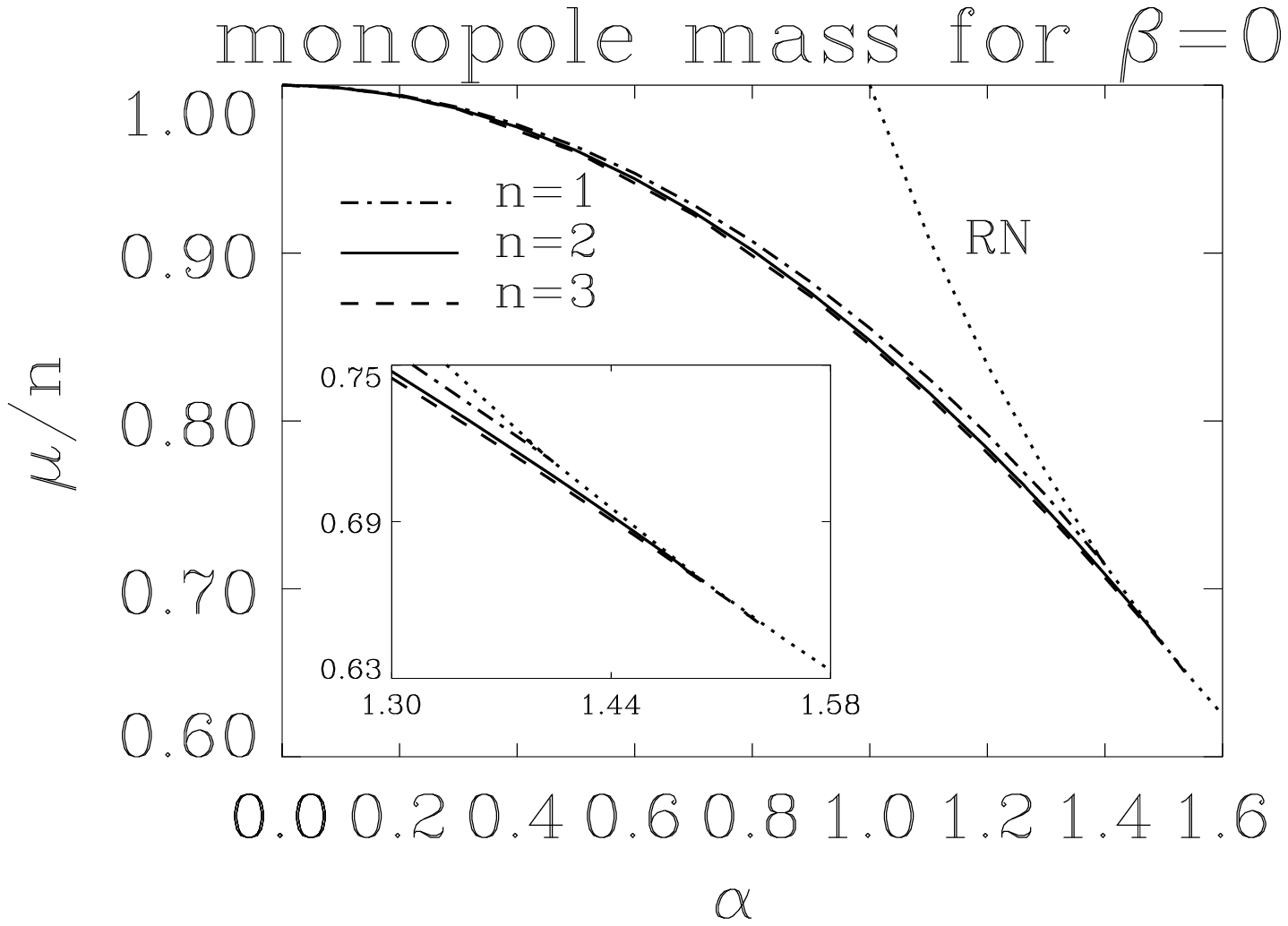}}
\caption{
The mass {\it per unit charge} in units of $4 \pi \eta /e$
is shown as a function of $\alpha$
for the $n=2$ and 3 multimonopoles in the BPS limit.
For comparison the mass of the monopole and the mass of the
extremal RN solutions are also shown.
}
\end{figure}

\begin{figure}\centering\epsfysize=8cm
\mbox{\epsffile{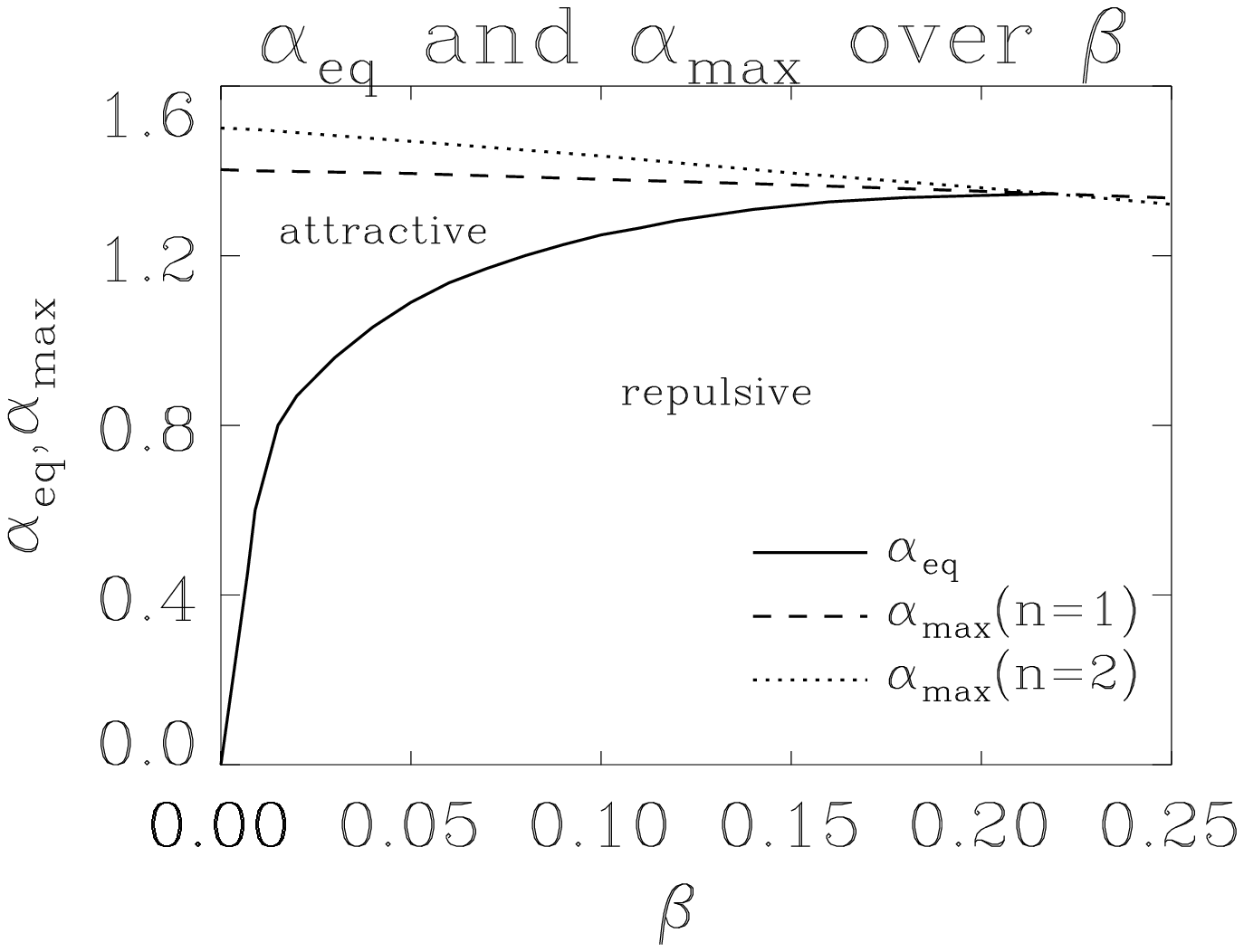}}
\caption{
The equilibrium value of $\alpha$, $\alpha_{\rm eq}$,
where the multimonopole mass {\it per unit charge} und monopole mass
equal one another, is shown as a function of $\beta$
for the $n=2$ multimonopoles.
Also shown are the maximal values $\alpha_{\rm max}$
for the $n=1$ and 2 (multi)monopoles.
}
\end{figure}


\begin{thebibliography}{000}
\bibitem{foot1}
 For $n=1$ monopoles
 for zero and very small Higgs self-coupling
 a short second branch extends backwards
 from the maximal value of $\alpha$, $\alpha_{\rm max}$,
 to the critical value of $\alpha$, $\alpha_{\rm cr}$,
 where the extremal RN solution is reached.
\bibitem{foot2}
 For $n=1$ monopoles
 for large Higgs self-coupling 
 the exterior solution is not of RN-type \cite{lw}.
\bibitem{foot3}
 The gauge invariant modulus of the Higgs field
 is very close to the Higgs field function $\phi_1$.
\end{thebibliography}
\end{document}